\documentclass[aps,prd,twocolumn,floatfix,preprintnumbers,nofootinbib,superscriptaddress]{revtex4}
\usepackage{mathtools}
\usepackage{graphicx}
\usepackage{siunitx}
\usepackage{color}
\usepackage{natbib}
\usepackage{multirow}
\usepackage{amsmath}

\newcommand{\dd}{\mathrm{d}}

\begin{document}
\preprint{LAPTH-014/21}
\title{Isotropic X-ray bound on Primordial Black Hole Dark Matter}
\author{J. Iguaz}
\email{iguaz@lapth.cnrs.fr}
\affiliation{Univ. Grenoble Alpes, Univ. Savoie Mont Blanc, CNRS, LAPTh, F-74940 Annecy, France}
\author{P. D. Serpico}
\email{serpico@lapth.cnrs.fr}
\affiliation{Univ. Grenoble Alpes, Univ. Savoie Mont Blanc, CNRS, LAPTh, F-74940 Annecy, France}
\author{T. Siegert}
\email{tsiegert@ucsd.edu}
\affiliation{Center for Astrophysics and Space Sciences, University of California, San Diego, 9500 Gilman Dr, 92093-0424, La Jolla, USA}
\date{\today}
\begin{abstract} 
We revisit the constraints on evaporating primordial black holes (PBHs) from the isotropic X-ray and soft gamma-ray background in the mass range $10^{16}-10^{18}$ g. We find that they are stronger than usually inferred due to two neglected effects: i) The contribution of the annihilation radiation due to positrons emitted in the evaporation process. ii) The high-latitude, Galactic contribution to the measured isotropic flux.  We study the dependence of the bounds from the datasets used, the positron annihilation conditions, and the inclusion of the astrophysical background.
We push the exclusion limit for  non-spinning PBH with monochromatic mass function as the totality of dark matter to 1.5$\times 10^{17}\,$g, which represents a $\sim$15$\%$ improvement with respect to earlier bounds and translates into almost an order of magnitude improvement in the PBH fraction in the already probed region. 
We also show that the inclusion of spin and/or an extended, log-normal mass function lead to tighter bounds. Our study suggests that the isotropic flux is an extremely promising target for future missions in improving the sensitivity to PBHs as candidates for dark matter.  
\end{abstract}

\maketitle
%%%%%%%%%%%%%%%%%%%%%%%%%%%%%%%%%%%%%%%%%%%%%%%%%%%%%%%%%%%%%%%%%%%%%%
\section{Introduction}
%%%%%%%%%%%%%%%%%%%%%%%%%%%%%%%%%%%%%%%%%%%%%%%%%%%%%%%%%%%%%%%%%%%%%%

Despite the overwhelming gravitational evidence for the dark matter (DM) phenomenon, the identification of its nature remains a formidable open problem in modern physics. None of the efforts to identify DM particles interacting weakly with standard model (SM) states, either via indirect observations or direct detection in shielded (often underground) detectors, have led to an unambiguous discovery.
While there is no shortage of DM particle candidates in extensions of the SM, there is no guarantee nor observational indication that the DM is made of microscopic fundamental particles.

Another established possibility (see e.g.\,\cite{Chapline:1975ojl}) is that DM is constituted by Primordial Black Holes (PBHs), hypothetical objects formed from the gravitational collapse of large overdensities in the early Universe~\cite{1967SvA....10..602Z,1974MNRAS.168..399C}. This has been recently revamped~\cite{Bird_2016,Sasaki_2016}
in the light of the LIGO-Virgo measurements of coalescing black hole binaries with typical masses of tens of M$_{\odot}$~\cite{Abbott_2016}.
Following studies (for a review, see~\cite{Carr_2020}) have concluded that PBHs in this mass window are constrained by a number of arguments to be no more than a sub-leading fraction of the whole DM content. 

However, there is also a consensus that PBHs lighter than about $10^{-10}\,M_{\odot}$ remain viable as DM, in principle down to masses where (the predicted, but yet unobserved) Hawking evaporation makes the PBHs cosmologically too short-lived, i.e. when $M\lesssim 5 \times 10^{14}$ g.
In practice, at least for masses up to $\sim 10^{17}$ g,  Hawking evaporation should be strong enough to induce signatures in cosmological observables (such as the Cosmic Microwave Background~\cite{Poulin_2017,St_cker_2018,PhysRevD.95.083006} and Big Bang Nucleosynthesis) and astrophysical ones (cosmic rays~\cite{Boudaud_2019}, gamma-rays and X-ray fluxes~\cite{DeRocco_2019,Laha_2019}, neutrinos~\cite{Dasgupta_2020}) to a level inconsistent with observations.

In this work we revisit the sensitivity of diffuse hard X-ray/soft gamma-ray flux measurements to PBHs in the mass range $10^{16}$ g -- $10^{18}$ g. These observations have been discussed in a number of works, the most recent ones  being~\cite{Laha_2020,Ballesteros:2019exr,Coogan:2020tuf,Carr_2016}.  
Compared with existing treatments of the {\it isotropic} flux (which is in particular the focus of Ref.~\cite{Ballesteros:2019exr}), we include two neglected effects: i) The photon contribution due to the annihilation radiation of positrons emitted in the evaporation process.
ii) The Galactic contribution to the isotropic flux: Since we are embedded inside the Galactic halo, as a result of line-of-sight integration, the Galactic flux, while peaking in the inner Galaxy, is non-zero in all directions.
This high-latitude Galactic contribution to the isotropic flux measured by the detectors cannot be separated from the truly extragalactic contribution (see Sec.\,\ref{sectionII}). However, it alters the expected spectral shape, a point we investigate in this work.
 As a result, we anticipate that the existing bounds too conservative.  We derive more realistic sets of bounds by comparing the computed PBH evaporation flux with the data compiled in~\cite{Ajello:2008xb}, also assessing the impact of the inclusion of models of the astrophysical background, one coming from the model of~\cite{Ueda_2014} and another one from the phenomenological fit adopted in~\cite{Ballesteros:2019exr}.

The paper is organized as follows: in Section \ref{sectionII} we provide all the computational details of the total isotropic flux, illustrate the data sets used, and the models considered to describe the astrophysical background. Then, in Section \ref{sectionIII} we discuss the statistical analysis followed to derive our limits on the abundance of PBHs and present our results (further analyses and considerations are reported in the Appendix). Finally, in Section \ref{sectionIV} we  present our conclusions and discuss future perspectives.

%%%%%%%%%%%%%%%%%%%%%%%%%%%%%%%%%%%%%%%%%%%%%%%%%%%%%%%%%%%%%%%%%%%%%%
\section{Datasets and Modeling}\label{sectionII}
%%%%%%%%%%%%%%%%%%%%%%%%%%%%%%%%%%%%%%%%%%%%%%%%%%%%%%%%%%%%%%%%%%%%%%

According to Hawking's groundbreaking results~\cite{1971MNRAS.152...75H}, a BH of mass $M_{\rm BH}$ radiates thermally with a temperature
\begin{equation}
    T_{\rm BH}=\frac{M^{2}_{\rm P}}{8\pi M_{\rm BH}},
    \label{tbh}
\end{equation}

\noindent where $M_{\rm P}$ is the Planck mass (natural units are adopted, unless indicated otherwise). A BH thus emits all kinematically accessible species $i$ with a blackbody distribution, modulo a species-dependent greybody factor $\Gamma_{i}(E,M_{\rm BH})$,
\begin{equation}
    \frac{\dd^2 N_{i}}{\dd E\dd t}=\frac{1}{2\pi}\frac{\Gamma_{i}(E,M_{\rm BH})}{e^{E/T_{\rm BH}}-(-1)^{2s}},
    \label{bhdist}
\end{equation}
where $s$ is the spin of the radiated particle and $E$ its energy. In order to compute the resulting spectra, we use the public software \texttt{BlackHawk}~\cite{Arbey:2019mbc}, which generates numerical files in the energy range of interest for all SM particles (also including secondary emission from fragmentation/decay). To compute the photon flux accounting for the effects we want to highlight, we focus on the photon and positron spectra.\\

We consider two main contributions to the total isotropic flux:
\begin{itemize}
    \item Hawking evaporation of PBHs into photons.
    \item Hard X-ray radiation produced by Hawking-radiated $e^+$s annihilating with  $e^-$s of the surrounding medium, typically via positronium (Ps) formation.
\end{itemize}
Each of these terms has both an {\it extragalactic} and a {\it Galactic} contribution. Below, we discuss all these contributions,  assuming a monochromatic mass distribution of non-rotating PBHs of mass $M$ accounting for a fraction $f_{\rm PBH}$ of the total DM density. In Sections~\ref{emf} and ~\ref{mrpbh} we will show and discuss what are the consequences of relaxing these assumptions.

\subsection{Direct photon component}
The direct extragalactic photon component due to the isotropically distributed contribution of all the extragalactic PBHs writes 
\begin{equation}
    \frac{\dd\phi^{\rm ext}_{\gamma}}{\dd E}=\frac{f_{\rm PBH}\,\Omega_{\rm DM}\rho_{c}}{4\pi M}\int^{z_{\rm max}}_{0}  \frac{\dd z}{H(z)}\frac{\dd^2 N_{\gamma}}{\dd E\dd t}(E(1+z))\,,
    \label{fluxeg}
\end{equation}

 \noindent with $f_{\rm PBH}$ the fraction of DM in the form of PBH, $H(z)=H_{0}\sqrt{\Omega_{\Lambda}+\Omega_{M} (1+z)^{3}}$,  $z_{\rm max} \sim O(500)$, $H_{0}=67.36$ km/(s$\,$Mpc), $\Omega_{\Lambda}=0.6847$, $\Omega_{\rm DM}=0.2645$ ~\cite{2020} and $\rho_{c}=9.1\times 10^{-30}$ $\text{g/cm}^{3}$. This is the only term, among the ones we consider, that has been included in~\cite{Ballesteros:2019exr}, for instance. Note that the integral is insensitive to the exact value of $z_{\rm max}$, provided that $z_{\rm max}\gg 10$. 
 
The {\it total} direct Galactic contribution $\Phi^{\rm gal}_\gamma$ depends on the integral along the line of sight of the DM distribution in the Galaxy, $\rho_g$, as
\begin{equation}
    \frac{\dd\Phi^{\rm gal}_\gamma(\hat \Omega)}{\dd E}=\frac{f_{\rm PBH}}{4\pi M} \frac{\dd^2 N_{\gamma}}{\dd E\dd t}{\cal D}(\hat \Omega)\,,
    \label{fluxgalanis}
\end{equation}
where $\hat \Omega$ denotes a direction in the sky and we defined as customary the {\it D-factor}
\begin{equation}
{\cal D}(\hat \Omega)\equiv \int_{\rm l.o.s.} \dd s\, \rho_{g}\,.
\end{equation}
We assume for $\rho_g$ a Navarro-Frenk-White profile~\cite{Navarro:1996gj} with parameters $r_{s}=9.98$ kpc, $\rho_{s}=2.2 \times 10^{-24}$ $\text{g/cm}^{3}$ extracted from a recent fit to Milky Way data~\cite{Karukes_2020}. The flux $\Phi^{\rm gal}_\gamma$ is clearly anisotropic, with a maximum towards the inner Galaxy, and as such it has been ignored in the calculation of isotropic fluxes in the past literature. However, it is finite  in {\it any} direction of the sky, since we are embedded in a DM (and thus, PBH) distribution which is not infinitesimally thin. As a result, observations sensitive to any residual flux of astrophysical origin measured over the solid angle $\Delta \Omega$ will also be sensitive to the {\it minimum} of this flux over $\Delta \Omega$. We have
\begin{equation}
\underset{\Delta \Omega}{\rm min}\,{\cal D}(\hat \Omega)\geq {\cal D}({\rm GAC})\equiv {\cal D}_{\rm min} \,,
\end{equation}
where GAC denotes the line of sight towards the Galactic anti-center, that is $l=\ang{180}$ and $b=\ang{0}$ in galactic coordinates, where the function ${\cal D}(\hat \Omega)$ attains its global minimum.
Thus, we conclude that each of the experiments is sensitive to a Galactic isotropic flux $\phi^{\rm gal}_\gamma$ bounded by  
\begin{equation}
    \frac{\dd\phi^{\rm gal}_\gamma}{\dd E}\geq  \left.\frac{\dd\phi^{\rm gal}_\gamma}{\dd E}\right|_{\rm min}\equiv \frac{f_{\rm PBH}}{4\pi M} \frac{\dd^2N_{\gamma}}{\dd E\dd t} {\cal D}_{\rm min}\,.
    \label{fluxgal}
\end{equation}

In conservative estimates below, we will adopt the right hand side of the
above equation as our proxy. A more realistic calculation, relying on the actual field of view used by each experiment to deduce their isotropic flux measurement, would yield a higher flux and thus more stringent constraints. Also, note that contrarily to analyses of the inner Galaxy data, the dependence on the chosen DM profile is very modest, since the signal depends on the density outside the solar circle, where it is well constrained by data.

\subsection{Photons from $e^+$ annihilations}
PBHs in the mass range considered also emit $e^\pm$ pairs. 
We include the extra photon component due to the annihilation
of the emitted positrons, under the sole assumption that steady state is attained. Although the details of the evolutionary energy track of a positron may be complex~\cite{Guessoum2005_511,Prantzos:2010wi}, this hypothesis is reasonable for the entire parameter space: For relativistic $e^\pm$, the energy loss timescale is comparable or faster than the Hubble expansion time even for conditions appropriate for the average cosmological medium at low redshift $z$~\cite{1975ApJ...196..689G}. For deeply non-relativistic $e^+$, atomic processes like excitation, ionization and charge-exchange become dominant. They are characterized by very large cross-sections, of the order of 10$^{-17}\,$cm$^2$ on hydrogen at keV energies~\cite{Guessoum2005_511}, again sufficient to ensure efficient cooling and eventual annihilation even on cosmologically diffuse medium at low $z$. The energy loss timescale is maximum around $\sim$MeV energies and at low-$z$. But at low-$z$ one must consider that the quasi-totality of DM, and hence PBHs, are bound in halos~\cite{2010MNRAS.401.1796A}. These are characterized by an overdensity of about 200 with respect to the cosmological density {\it at formation epoch}. A fortiori, then, we can assume that the byproducts of PBH evaporation encounter a medium which is at least two to three orders of magnitude denser than the cosmological one. This is enough to ensure that the typical loss timescale for atomic or plasma losses $(\sigma_T\,n_e)^{-1}$ is shorter than the age of the universe, with $\sigma_T$ the Thompson cross section and $n_e$ the electron density.  At high-$z$, the inverse Compton loss timescale on the CMB is much faster than the Hubble time even in the diffuse cosmological medium. 

In the following, we will not consider the uncertain photon emission via $e^\pm$ energy loss byproducts. Although part of these may end up in the energy region of interest, typically they are less energetic and we shall discard them, also because the exact calculation relies on a number of assumptions on the relative importance of different energy loss channels. On the other hand, a quasi-steady state $e^+$ annihilation radiation is unavoidable under physically realistic assumptions, and will fall in the X-ray band. 
Formally, we can write these contributions similarly to Eq.~(\ref{fluxeg}) and Eq.~(\ref{fluxgal}), respectively, as 
\begin{equation}
    \frac{\dd\phi^{\rm ext}_{e^+}}{\dd E}=\frac{f_{\rm PBH}\Omega_{\rm DM}\rho_{c}}{4\pi M}\int^{z_{\rm max}}_{0}  \frac{\dd z}{H(z)}\frac{\dd^2 N_{\gamma}^{\rm ann.}}{\dd E\dd t}(E(1+z))\,,
    \label{fluxegPOS}
\end{equation}
and
\begin{equation}
    \frac{\dd\phi^{\rm gal}_{e^+}}{\dd E}\simeq \frac{f_{\rm PBH}}{4\pi M} \frac{\dd^2 N_{\gamma}^{\rm ann.}(E)}{\dd E\dd t} {\cal G}_{\rm min}\,.\label{fluxgalPOS}
\end{equation}
Here, $\dd^2 N_{\gamma}^{\rm ann.}(E)/\dd E\dd t$ is the energy-differential spectrum of photons per annihilation, times the number of annihilations per unit time, while 
${\cal G}_{\rm min}$ is analogous to the ${\cal D}_{\rm min}$ factor, but now taking into account the density of the positrons in the Galaxy at annihilation (steady state). We expect that, due to diffusion smoothing out gradients in the emission, ${\cal G}_{\rm min}> {\cal D}_{\rm min}$. Nonetheless, our  results will be presented with the conservative choice ${\cal G}_{\rm min}={\cal D}_{\rm min}$.

To make these fluxes explicit, we must know how photons are produced in annihilations.
If all of them were coming from direct $e^\pm$ annihilation, that is when the fraction of positronium formed vanishes or $f_{\rm Ps}=0$, each $e^+$ would lead to two photons emitted with energy $511$ keV (neglecting the annihilating particles kinetic energy, expected to be small). As a result, in this extreme case, the Galactic contribution writes 
\begin{equation}
   \frac{\dd\phi^{\rm gal}_{0}}{\dd E}=\frac{f_{\rm PBH}\, {\cal G}_{\rm min}}{4\pi M}2\,\delta(E-m_e)Y_{e^+}\,\Gamma_{\rm PBH},
    \label{511gal}
\end{equation}
where we defined
\begin{equation}
 Y_{e^+}\,\Gamma_{\rm PBH}\equiv\int\dd E'\frac{\dd^2 N_{e^{+}}}{\dd E'\dd t}\,.
    \label{yield}
\end{equation}
This gives the PBH evaporation rate $\Gamma_{\rm PBH}$ (essentially constant over cosmological timescales, for the masses considered) times the number of $e^+$ emitted, $ Y_{e^+}$, and corresponds to the integral over energy of the output of the \texttt{BlackHawk} software divided by 2.
In numerical evaluations, the spectral line at $m_e$ in Eq.\,(\ref{511gal}) is modelled as a  Gaussian with $\sigma=1$ keV. This is motivated by the expectation that most positrons would either form Ps after thermalization ($\sim 1.2\,\mathrm{keV}$) or via radiative recombination with free electrons ($\sim 1.0\,\mathrm{keV}$) \citep{Guessoum2005_511}.

The corresponding redshifted contribution to the extragalactic flux writes
\begin{equation}
   \frac{\dd\phi^{\rm ext}_{0}}{\dd E}=\frac{f_{\rm PBH}\Omega_{\rm DM}\rho_{c}}{4\pi M}\frac{2\,\theta(m_e-E)}{H(m_{e}/E-1)}  Y_{e^+}\,\Gamma_{\rm PBH}\,,
    \label{511eg}
\end{equation}
where  we denoted by $\theta$ the Heaviside step-function.

In the opposite limit where all annihilations proceed via Ps formation, i.e. $f_{\rm Ps}=1$, the spectrum will depend on the spin state: para-Ps will yield two photons of $511$ keV as above, while ortho-Ps annihilation will yield three photons with an energy distribution $h_{3\gamma}$($E/m_{e}$) \cite{1949PhRv...75.1696O}, following a ratio 2:9 set by quantum statistics. Indeed,  Ps is formed in a singlet (para-Ps) or a triplet (ortho-Ps) state in a ratio 1:3 due to the number of spin states $2s+1$, and the ratio of photons emitted in para-Ps over ortho-Ps is 2/3, hence the aforementioned ratio. Therefore, in this scenario the Galactic isotropic flux will write as
\begin{equation}
\begin{split}
   \frac{\dd\phi^{\rm gal}_{1}}{\dd E}=&\frac{f_{\rm PBH}\, {\cal G}_{\rm min}}{4\pi M}Y_{e^+}\,\Gamma_{\rm PBH}
    \\&\times \left(2\,N_{p}^{\rm gal}\delta(E-m_e)+3\,N_{o}^{\rm gal}\, \frac{h_{3\gamma(E/m_{e})}}{E}\right),
    \label{Posgal}
\end{split}
\end{equation}
and the extragalactic one as
\begin{equation}
\begin{split}
   & \frac{\dd\phi^{\rm ext}_{1}}{\dd E}= \frac{f_{\rm PBH}\Omega_{\rm DM}\rho_{c}}{4\pi M}  Y_{e^+}\,\Gamma_{\rm PBH}\left(2\,N^{\rm ext}_{o}\frac{\theta(m_e-E)}{H(m_{e}/E-1)}\right.\\
   &\left.+3N^{\rm ext}_{p}\int_0^{z_{\rm max}} \dd z\frac{f_{3\gamma}[E(1+z)/m_{e}]}{H(z)}\right)
    \label{Poseg}
\end{split}
\end{equation}
where the first term on the right hand side of Eqs.\,(\ref{Posgal},\ref{Poseg}) corresponds to para-Ps formation and the second one to ortho-Ps formation. We introduce the factors $N^{i}_{p}$ and $N^{i}_{o}$ to normalize to the same value for the integrated energy flux as in the case $f_{\rm Ps}=0$.

It has been shown by \citet[][and refereces therein]{Guessoum2005_511}, that for typical conditions of the neutral warm ISM,  $f_{\rm Ps}\gtrsim 0.8$ for He, and $f_{\rm PS}>0.95$ for H. This Ps formation `in flight' decreases roughly inversely  with the ionisation fraction. The cross sections for direct annihilation with free and bound $e^-$s are several orders of magnitude smaller than the charge exchange reactions required to build Ps, and become only important for temperatures below $\sim 8000\,\mathrm{K}$. For these lower temperatures, the radiative recombination with free $e^-$s is the dominant channel. It can thus be expected that most $e^+$s ejected in the halo are prone to form Ps, which makes $f_{\rm Ps} \sim 1.0$ a more realistic case for Milky Way-like conditions. For the enhanced densities at higher redshifts, the arguments are similar. As a result, we consider 
$f_{\rm Ps}= 1.0$ our benchmark choice, but we shall also discuss the impact of the opposite limit $f_{\rm Ps}= 0$, which provides a very conservative bracketing of the spectral uncertainty.

The total isotropic differential flux due to PBH is summarized as:
\begin{equation}
 \frac{\dd\phi_{\rm PBH}}{\dd E} =\frac{\dd\phi^{\rm gal}_\gamma}{\dd E}+\frac{\dd\phi^{\rm ext}_\gamma}{\dd E}+
    \begin{cases}
      \frac{\dd\phi^{\rm gal}_{0}}{\dd E}+\frac{\dd\phi^{\rm ext}_{0}}{\dd E} & \text{if $f_{\rm Ps}=0$}\\
      & \\
     \frac{\dd\phi^{\rm gal}_{1}}{\dd E}+\frac{\dd\phi^{\rm ext}_{1}}{\dd E} & \text{if $f_{\rm Ps}=1$}
    \end{cases}    
    \label{totalisoflux}
\end{equation}

\begin{figure}[htbp]
  \centering
    \includegraphics[width=0.45\textwidth]{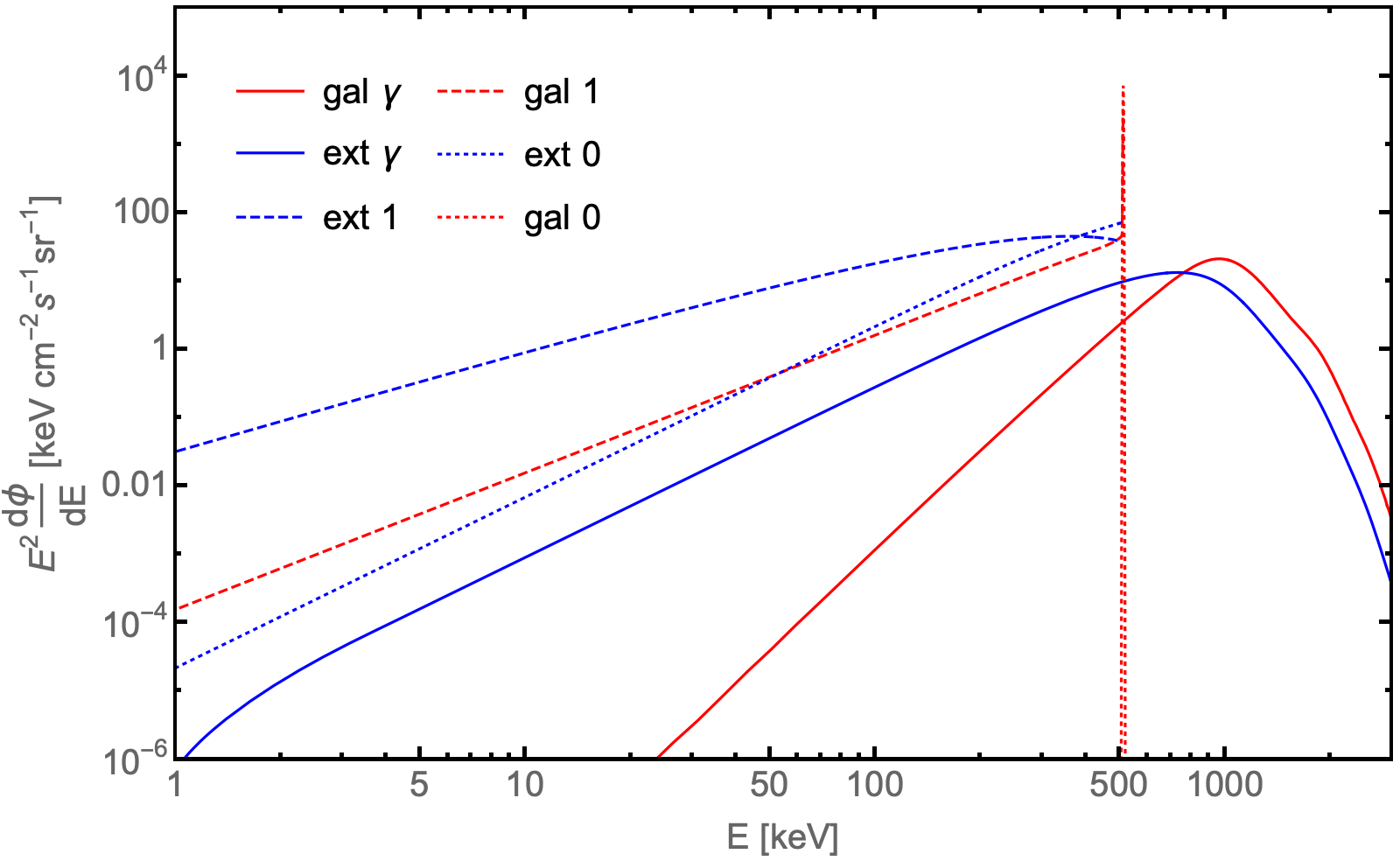}
    \caption{Different contributions to the total isotropic differential flux, for $M_{\rm PBH}=7\times 10^{16}$g.}
    \label{isoflux}
\end{figure}

\begin{figure}[htbp!]
  \centering
    \includegraphics[width=0.45\textwidth]{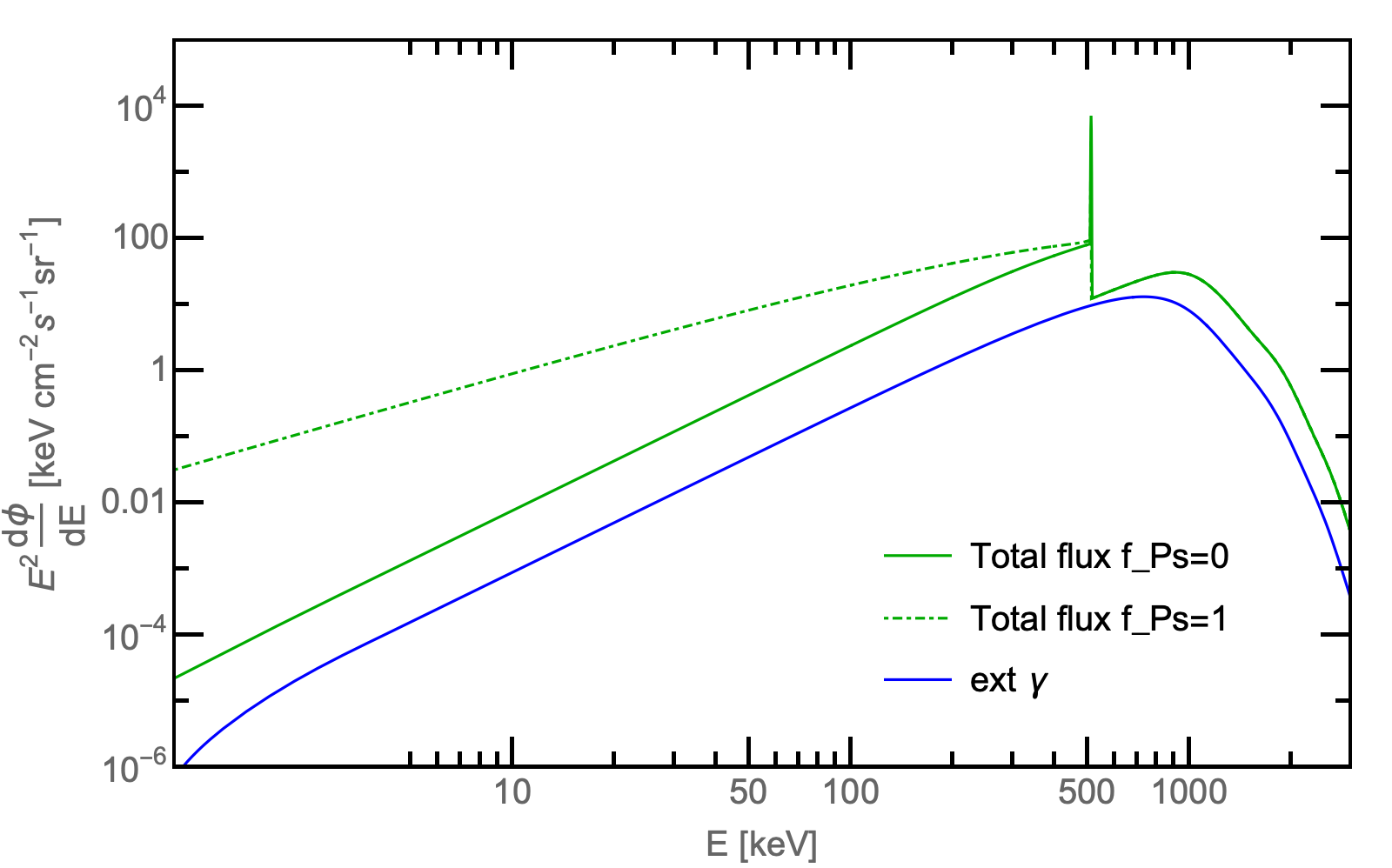}
    \caption{The extragalactic flux typically considered in the literature compared to the total isotropic differential flux for $f_{\rm Ps}=0$ and $f_{\rm Ps}=1$, for $M_{\rm PBH}=7\times 10^{16}$g.}
    \label{totalflux}
\end{figure}

In Figure \ref{isoflux} we show the different contributions to the isotropic photon flux of Eq.~(\ref{totalisoflux}) and in Figure~\ref{totalflux} the total isotropic flux for a representative value of $M_{\rm PBH}=7\times 10^{16}\,$g. It is clear that, independently by any residual uncertainty on the value of $f_{\rm Ps}$, the  extragalactic flux typically considered in the literature (solid blue line) is significantly lower, possibly by more than one order of magnitude, than the actual total flux predicted. \\

\begin{figure}[htbp]
  \centering
    \includegraphics[width=0.45\textwidth]{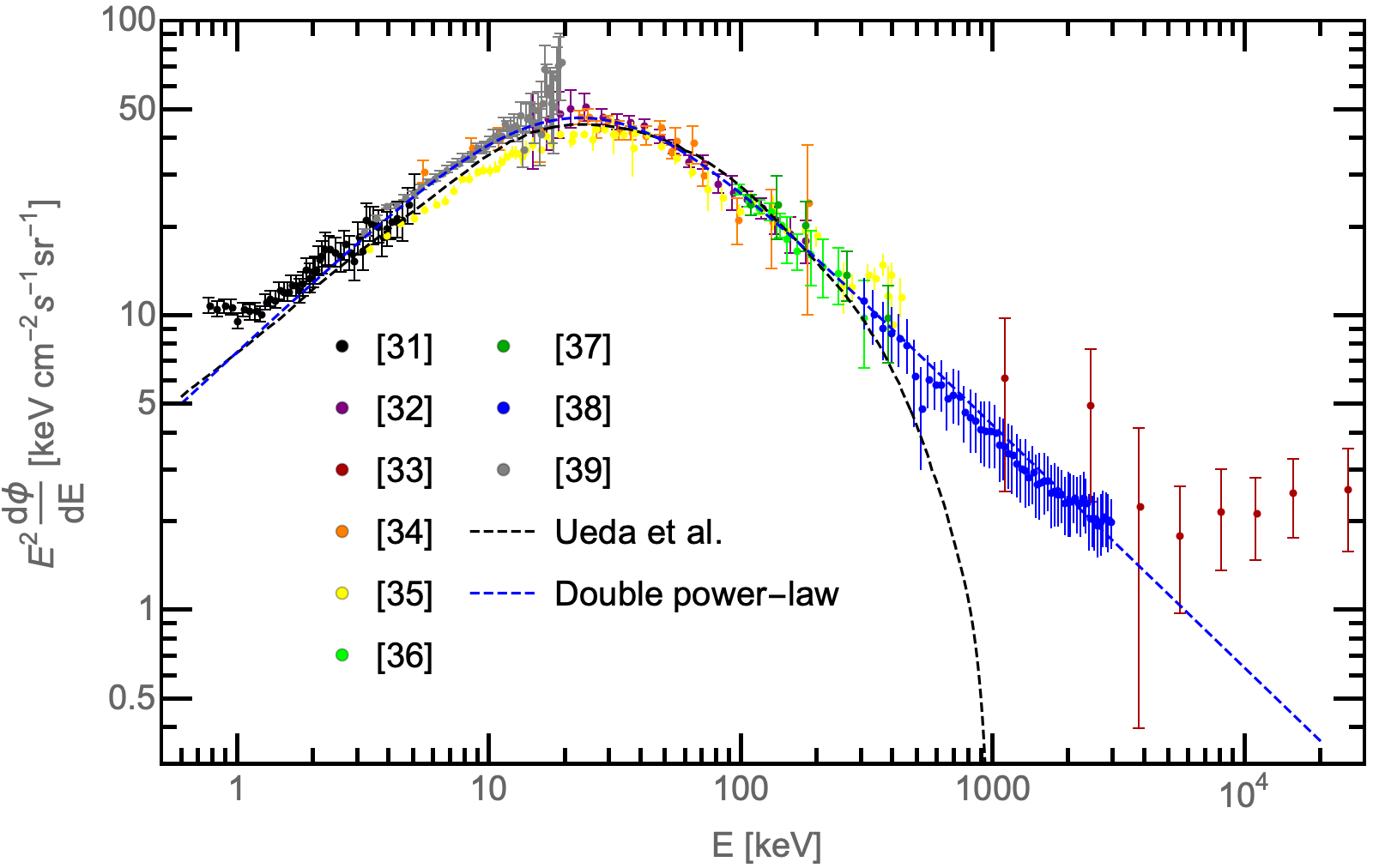}
    \caption{Cosmic X-ray background as measured by different experiments, Ueda model~\cite{Ueda_2014} (black dashed line) and a double power-law fit~\cite{Ballesteros:2019exr} (blue dashed line).}
    \label{datamodel}
\end{figure}

In order to set bounds on $f_{\rm PBH}$, we compare the differential flux in Eq.~(\ref{totalisoflux}) with the available data in the range $\sim$1 keV to 25 MeV observed by ASCA~\cite{1994PASJ...46L..37T}, Swift/BAT~\cite{2005SSRv..120..143B}, Comptel~\cite{2000AIPC..510..467W}, INTEGRAL~\cite{Churazov2007A&A...467..529C}, HEAO-1~\cite{1999ApJ...520..124G}, HEAO-A4~\cite{1997ApJ...475..361K}, Nagoya~\cite{1975Ap&SS..32L...1F}, SMM~\cite{1997AIPC..410.1223W} and RXTE~\cite{2003A&A...411..329R},
as compiled in~\cite{Ajello:2008xb}. Whenever no explicit error on energy is provided, we have considered the flux to refer to the bin defined by the half-difference of neighbouring energies. 
Additionally, it is clear that the bulk of the  X-ray and gamma-ray background {\it cannot} be attributed to PBHs, since the unresolved counterparts of well-known objects (such as active galactic nuclei) are expected to largely account for the observations. To illustrate that,  we employ the population synthesis model of active galactic nuclei proposed by Ueda et al. in~\cite{Ueda_2014}, which provides a reasonable match to the observations up to $\sim$200 keV (black dashed line in Figure \ref{datamodel}). We also display the double power-law fit of the background employed in ref.~\cite{Ballesteros:2019exr}, which is phenomenologically based but reproduces data in a larger dynamic range.

%%%%%%%%%%%%%%%%%%%%%%%%%%%%%%%%%%%%%%%%%%%%%%%%%%%%%%%%%%%%%%%%%%%%%%
\section{Results}\label{sectionIII}
%%%%%%%%%%%%%%%%%%%%%%%%%%%%%%%%%%%%%%%%%%%%%%%%%%%%%%%%%%%%%%%%%%%%%%

\subsection{Without background modelling}

\begin{figure}[htbp!]
  \centering
    \includegraphics[width=0.45\textwidth]{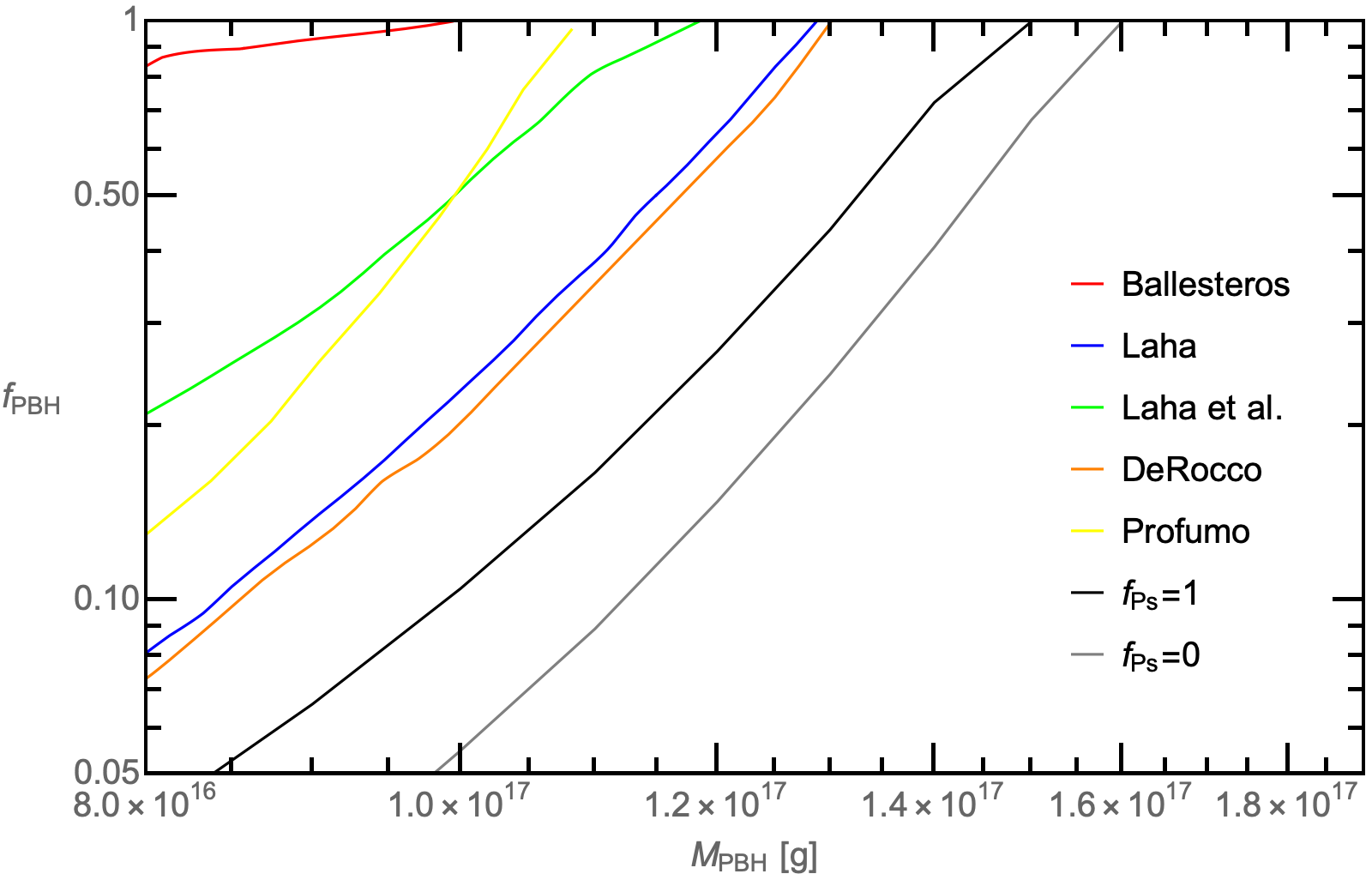}
    \caption{Bounds derived without background modelling for $f_{\rm Ps}=0$ (gray) and $f_{\rm Ps}=1$ (black). For comparison, we also display other bounds in the literature:\cite{Ballesteros:2019exr} in red, \cite{Laha_2019} in blue, \cite{Laha_2020} in green, \cite{DeRocco_2019} in orange and \cite{Coogan:2020tuf} in yellow.}
    \label{boundscombinedallnew}
\end{figure}

We derive different sets of bounds on the allowed PBH abundance. The most conservative
one, similar to what has been done e.g. in~\cite{Laha_2020,Ballesteros:2019exr,Coogan:2020tuf}, requires that the flux solely due to PBH, Eq.~(\ref{totalisoflux}), does not exceed any measurement data point by more than 2 $\sigma$.  In Fig.~\ref{boundscombinedallnew} we report the bounds thus obtained,  for $f_{\rm Ps}=1$ (black/darker)  or $f_{\rm Ps}=0$ (gray/lighter), compared with others published in the literature, zooming on the most interesting parameter space for PBH DM, around 10$^{17}\,$g. The stronger constraints for $f_{\rm Ps}=0$ depend on the tighter SMM bounds obtained when the prominent line at 511 keV is present. The impact of these new bounds can be quantified via two alternative metrics: the improvement on the constrained $f_{\rm PBH}$, in the region of masses already probed for the totality of DM, or the extra parameter space now excluded as the totality of DM in terms of PBH mass. In the second case, we rule out the totality of DM in the form of PBHs for values of $M_{\rm PBH}$ lower than $1.5-1.6 \times 10^{17}\,$g i.e. extending by about 20\% the excluded range of $M_{\rm PBH}$. Adopting the first metric, the improvement over the bounds reported in~\cite{Ballesteros:2019exr} (red line) appears even more substantial; in particular at $M\lesssim 10^{17}\,$g, we now constrain $f_{\rm PBH}\ll 1$, i.e. by more than one order of magnitude. 
Presently derived bounds are also more stringent than those coming from analyses of other observables. Compared to several of those, relying on inner Galaxy fluxes, we also remind the reader that our constrains are
poorly sensitive to the details of the inner DM halo profile.

In Appendix~\ref{appC}, for the purpose of comparing directly our results with those of Ref.~\cite{Ballesteros:2019exr}, we also present the results of adopting their ``conservative'' method, which extends this simple and robust criterion to a multi-bin analysis. In  the appendix~\ref{appC} we also display the bounds obtained for each dataset individually, and discuss how the bounds are affected by eliminating the most constraining dataset. 

\subsection{With background modelling}\label{wbm}

\begin{figure}[htbp]
  \centering
    \includegraphics[width=0.45\textwidth]{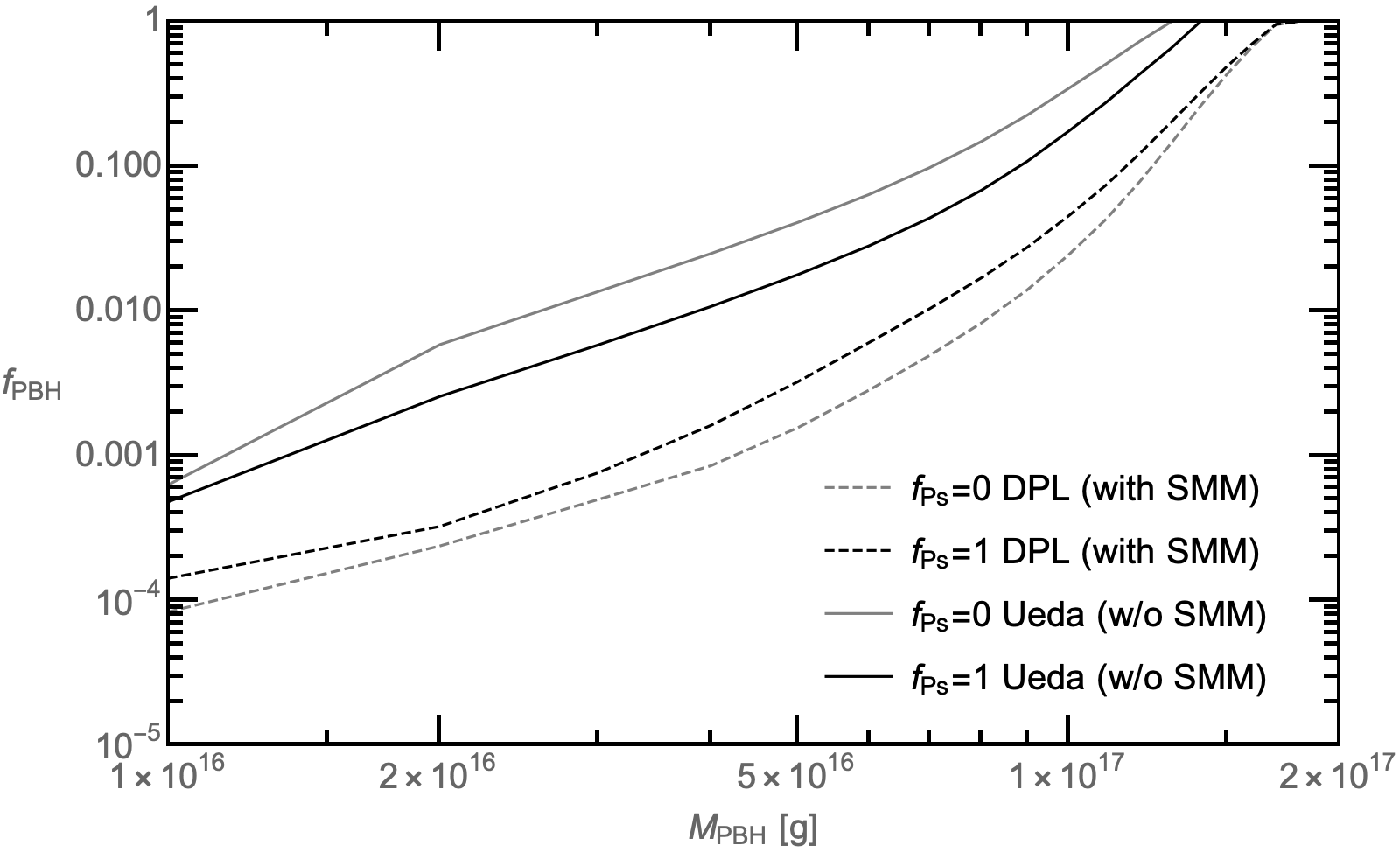}
    \caption{Bounds derived with background modelling according to Ueda et al. (solid) and a double power-law (dashed) for $f_{\rm Ps}=0$ (gray) and $f_{\rm Ps}=1$ (black).}
    \label{others}
\end{figure}

A more realistic bound takes into account that the flux from PBH is a poor representation of the diffuse photon measurements, and that most of the data must be accounted by astrophysical processes. As data from the different instruments are approximately normal distributed, we calculate the $\chi^2$ as the logarithm of the normal likelihood:
\begin{equation}
  \chi^{2}=\sum^{N_{T}}_{i}\frac{\left(D_{i}-A_{i}\right)^{2}}{\Delta^{2}_{i}}\,,
    \label{estimatorBG}
\end{equation}
summing over all energy bins the squared difference between the datapoints $D_i$ and the model $A_i$, normalized to the
data uncertainties $\Delta_i$ (assumed uncorrelated). The 95 $\%$ C.L. bound on $f_{\rm PBH}$ is obtained by computing for each PBH mass $M$: i) The value $f_{*}$ of $f_{\rm PBH}$ for which $\chi^{2}$ is minimized; ii) the value $f_{\rm max}$ such as $\chi^{2}(f_{\rm max})-\chi^{2}(f_*)= 3.84$. The model $A_i$ includes both the astrophysical background and the PBH flux. Since the model~\cite{Ueda_2014} is not suitable to explain the data above about 200 keV, we perform two types of analyses: a) using model~\cite{Ueda_2014} but excluding SMM and Comptel data; b) adopting 
the double power-law fit from ref.~\cite{Ballesteros:2019exr} and excluding Comptel data.
The resulting bounds are reported in Fig.~\ref{others}. While the bounds in the case a) (solid lines) are comparable with the ``no background'' analysis, the ones for case b) (dashed lines) lead to a factor 3 to 15 tighter constraints, pushing the limiting PBH mass for constituting the totality of the DM to $M=1.7\times 10^{17}\,$g. 

Although case b) cannot be considered a tight constraint, since the the double power-law background from ref.~\cite{Ballesteros:2019exr} is purely phenomenological and lacks a deep astrophysical motivation, it illustrates the power of current data in improving the sensitivity to PBHs. This certainly provides further motivation towards an improved understanding of the diffuse flux.

\subsection{Extended mass function}\label{emf}

One may wonder how the previously derived bounds on $f_{\rm PBH}$ change if relaxing somewhat the monochromatic mass function hypothesis. To quantitatively address this question, we consider a lognormal mass function $\psi(M)$ of the form

\begin{equation}
    \psi(M)=\frac{f_{\rm PBH}}{\sqrt{2\pi}\sigma M}\exp \left( -\frac{\log^{2}(M/M_{c})}{2\sigma^{2}} \right),
\end{equation}
where $M_{c}$ corresponds to the peak of the distribution and $\sigma$ parameterizes its width. In the past, this form or similar ones have been used to reproduce pretty well some theoretically predicted PBH mass functions, as shown in~\cite{Green:2016xgy} and reviewed in~\cite{Carr_2017}. 
In~\cite{Gow:2020bzo}, values of $\sigma\simeq 1/3\div 1/2$ have been argued to correspond to more realistic descriptions of the mass function than the monochromatic approximation.

\begin{figure}[htbp]
  \centering
    \includegraphics[width=0.45\textwidth]{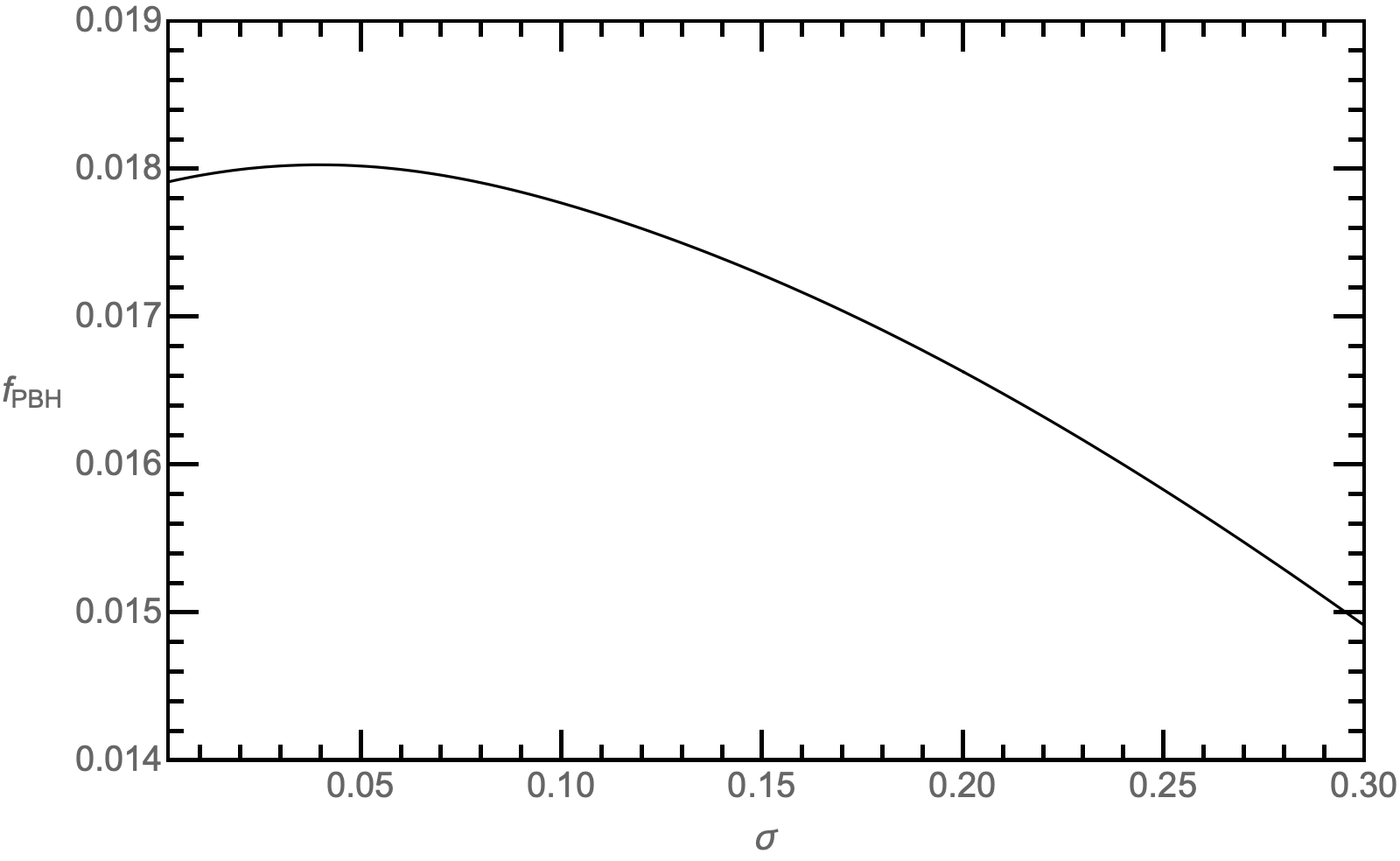}
    \caption{Maximum value for $f_{\rm PBH}$ as a function of the width of the mass distribution $\sigma$ for $M_{c}=5\times 10^{16}$ g and $f_{\rm Ps}=1$ and background modelling following ref.\,\cite{Ueda_2014} .}
    \label{extended}
\end{figure}

It turns out that, as long as the corresponding monochromatic mass bounds on $f_{\rm PBH}$ in a range of masses $\Delta M$ are known, the constraint on an extended mass function can be obtained as a byproduct, following the procedure described in~\cite{Carr_2017} (in particular, we apply their equation 12). In Figure~\ref{extended}, we show the results for the particular case in the middle of our parameter space $M_{c}=5\times 10^{16}$ g with background modelling and $f_{\rm Ps}=1$ (black solid line in Figure \ref{boundscombinedallnew}).

We find that the bound on $f_{\rm PBH}$ is rather insensitive to $\sigma$ as long this is below some value (here, about 0.1), while tightening above. Thus, at least as long as the mass support of the distribution is not too large (a decade in mass or so), we expect the results previously obtained for the monochromatic case to err, if anything, on the conservative side.
This is a rather generic conclusion for a number of constraints, as discussed in~\cite{Carr_2017}.

%%%%%%%%%%%
\subsection{Rotating PBHs}\label{mrpbh}
\begin{figure}[htbp]
  \centering
    \includegraphics[width=0.45\textwidth]{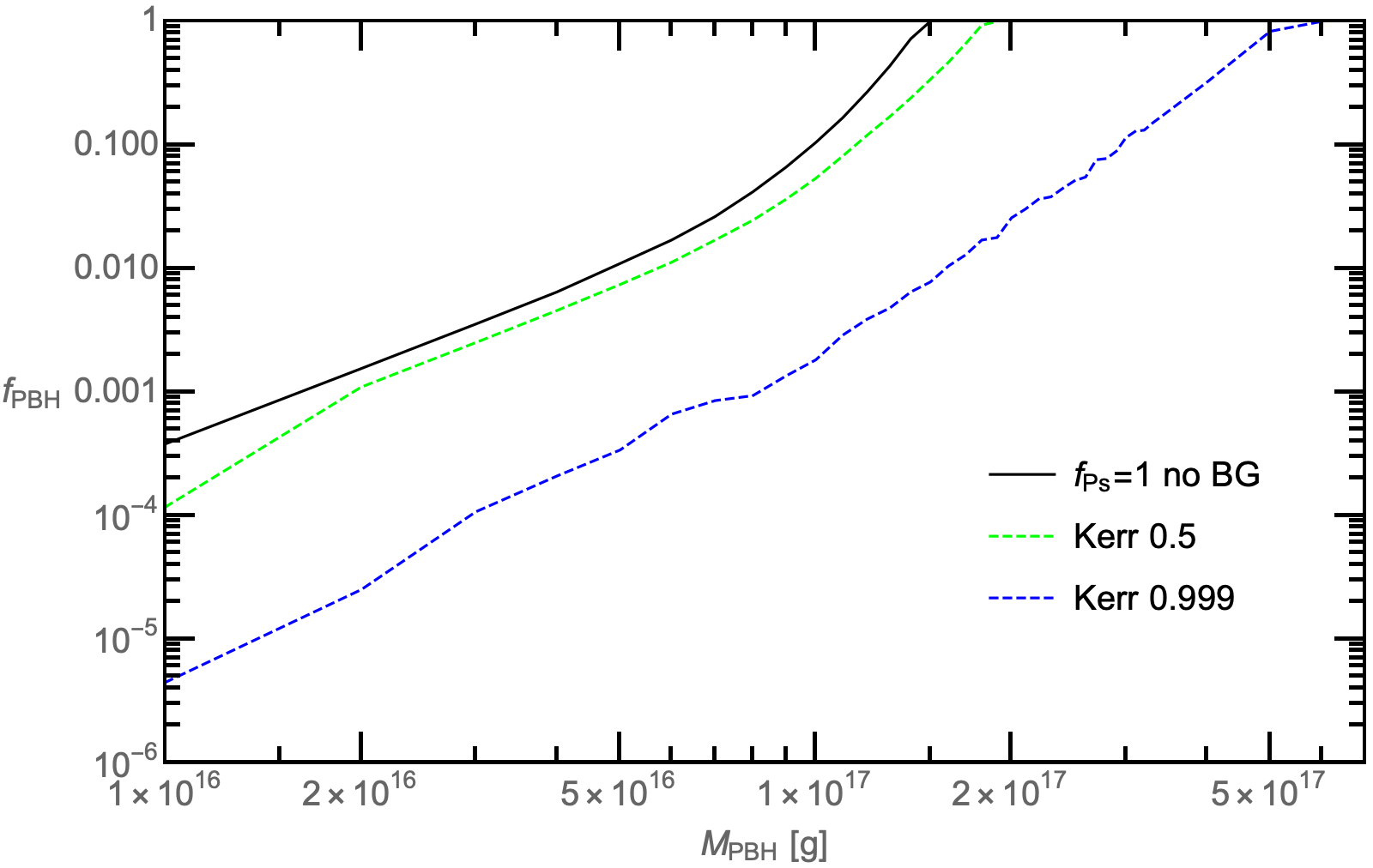}
    \caption{Bound derived for nonrotating (solid black), mildly rotating (dashed green) and close-to-maximally rotating (dashed blue) PBHs for $f_{\rm Ps}=1$ and no background modelling.}
    \label{boundKerr}
\end{figure}

%%%%%%%%%%%
In the previous sections we considered the simplest case of non-rotating PBHs. This is a reasonable proxy for PBHs at formation, whose expected dimensionless spin parameter is $a\lesssim 0.01$~\cite{DeLuca:2019buf}. However, since larger values of spin could also result  from dynamical evolution processes (like mergers of comparable-mass, non-spinning BH which produce a peak at $a\sim 0.7$~\cite{Berti:2008af}), it is interesting to examine how sensitive the bounds are to non-vanishing spin values. The most extreme case is to consider maximally rotating PBHs: We proceed in a similar fashion as in the monochromatic mass function case treated before, but for the intrinsic spin set close to maximum $a=0.999$ when generating the Hawking spectra with the \texttt{BlackHawk} software. 
We show our results in Figure \ref{boundKerr}, where we compare them with the corresponding non-rotating case, as well as with a rotating case with $a=0.5$. Constraints improve only mildly for moderately rotating PBHs, but are significantly enhanced for close to maximally rotating PBHs. In the latter case, they extend to a factor 4 higher mass or, at a given PBH mass, they are almost hundred times stronger. Similar improvements are typically present for different data combinations and background assumptions.  These results allow us to conclude that the constraints presented in the paper are therefore conservative.

%%%%%%%%%%%%%%%%%%%%%%%%%%%%%%%%%%%%%%%%%%%%%%%%%%%%%%%%%%%%%%%%%%%%%%
\section{Discussion and conclusions}\label{sectionIV}
%%%%%%%%%%%%%%%%%%%%%%%%%%%%%%%%%%%%%%%%%%%%%%%%%%%%%%%%%%%%%%%%%%%%%%

In this paper we have considered a population of PBHs constituting all or part of the dark matter (DM),  with a monochromatic mass distribution in the range $10^{16}\,{\rm g} - 10^{18}\,$g. We have set new upper bounds on the abundance of such objects by comparing existing experimental data with the isotropic photon flux from Hawking evaporation, coming both from photons radiated directly in the hard X-ray/soft gamma ray band and from positron annihilation radiation.
We have considered both the extragalactic and the Galactic halo contributions, and assessed the dependence from the channel through which positrons annihilate: Either directly with interstellar electrons, yielding a line at $511$ keV, or where all positrons form positronium, which annihilates either in two photons or three photons. Overall, we have shown that the total flux can be much larger than the sole extragalactic photon component typically considered in the literature.

Both conservative bounds without any background assumption and more realistic bounds accounting for the astrophysical background model are  stronger than other bounds published in the literature, in particular when $f_{\rm PBH}\sim 1$.  We exclude that the totality of DM is in the form of PBHs for values of $M_{\rm PBH}$ lower than$1.5-1.6 \times 10^{17}\,$g.
These bounds on $f_{\rm PBH}$ become more stringent if the mass function is extended (e.g. by about 20\% if having a lognormal width of $\sigma=0.3$) or if PBHs have a non-negligible spin, either of primordial origin or dynamically acquired (up to two orders of magnitude for the spin parameter $a=1$).  

A number of improvements are possible. Our analysis could be extended to lighter PBHs,  where bounds are already restricting $f_{\rm PBH}\ll 1$, and account for final state radiation of $e^\pm, \mu^\pm, \pi$'s, as discussed in~\cite{Coogan:2020tuf}. 
For the most interesting region above $10^{17}\,$g, we expect the bounds to improve significantly if better modeling of the diffuse flux were available, as illustrated in Sec.~\ref{wbm} and, above all, thanks to future observations in the X-ray or soft gamma-ray band, as already stressed in\,\cite{Ballesteros:2019exr}. These experiments have a rich astrophysics case, ranging from multimessenger searches of counterparts of gravitational wave or neutrino events, to shedding light on the acceleration mechanisms in Galactic and extragalactic objects, to nucleosynthesis studies via line spectroscopy. Their relevance also for PBH searches has been recently and extensively treated in the dedicated paper\,\cite{Ray:2021mxu}, so we will not repeat it here. However, we want to emphasize that it is not only important to reduce error bars on future diffuse measurements, but also resolve a significant part of this flux into astrophysical sources. The combined advances along these directions will likely allow one to extend the sensitivity to PBH DM by one order of magnitude in mass at least\,\cite{Ray:2021mxu}. Hopefully, in the coming decade one may either close the remaining viable gap in the PBH DM parameter space, or perhaps bump into a surprising discovery.

\begin{acknowledgments}
We thank G. Ballesteros for comments on an earlier version of this article, and M. Korwar and S. Profumo for pointing out a misinterpretation of the positron spectrum output of BlackHawk.
JI is supported by a CNRS International collaboration program. This work is partially supported by the project {\it MAGraW} under the program ``Initiatives  de  Recherche Strat\'egique'' - IDEX Univ. Grenoble-Alpes (P.I.: PDS). Thomas Siegert is supported by the German Research Foundation (DFG-Forschungsstipendium SI 2502/1-1 \& 2502/3-1)
\end{acknowledgments}
%%%%%%%%%%%%%%%%%%%%%%%%%%%%%%%%%%%%%%%%%%%%%%%%%%%%%%%%%%%%%%%%%%%%%%
\bibliography{biblio}
%%%%%%%%%%%%%%%%%%%%%%%%%%%%%%%%%%%%%%%%%%%%%%%%%%%%%%%%%%%%%%%%%%%%%%
\appendix

%%%%%%%%%%%%%%%%%%%%%%%%%%%%%%%%%%%%%%%%%%%%%%%%%%%%%%%%%%%%%%%%%%%%%%
\newpage
\section{Extended study of the bounds}\label{appC}
%%%%%%%%%%%%%%%%%%%%%%%%%%%%%%%%%%%%%%%%%%%%%%%%%%%%%%%%%%%%%%%%%%%%%%
Here, we present the exclusion limits obtained with further analyses, and discuss the outcomes of some sanity checks on the bounds.

\begin{figure}[htbp!]
  \centering
    \includegraphics[width=0.45\textwidth]{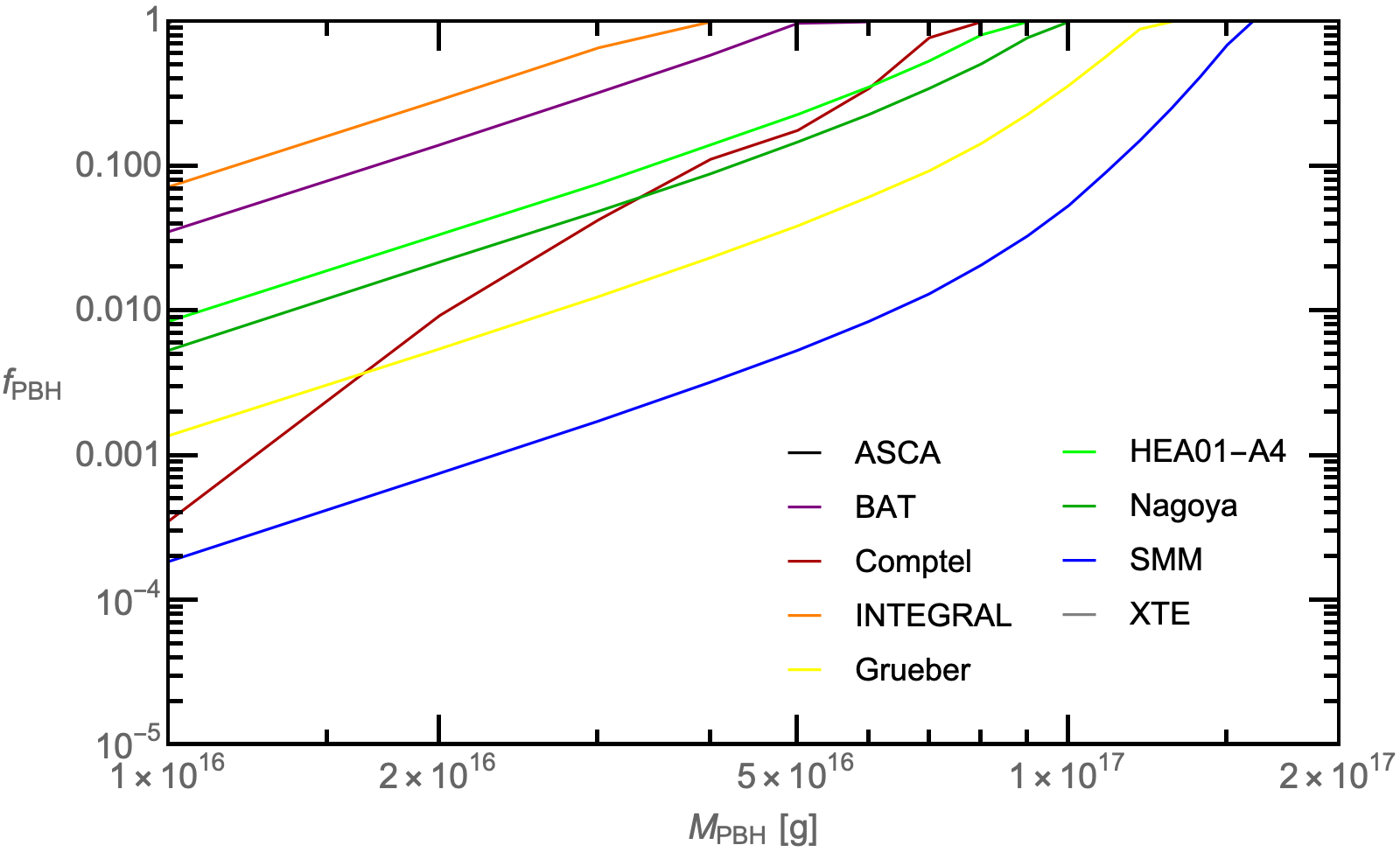}
    \caption{Bounds with no background modelling and $f_{\rm Ps}=0$ for the considered datasets.}
    \label{boundsfPs0noBG}
\end{figure}

\begin{figure}[htbp!]
  \centering
    \includegraphics[width=0.45\textwidth]{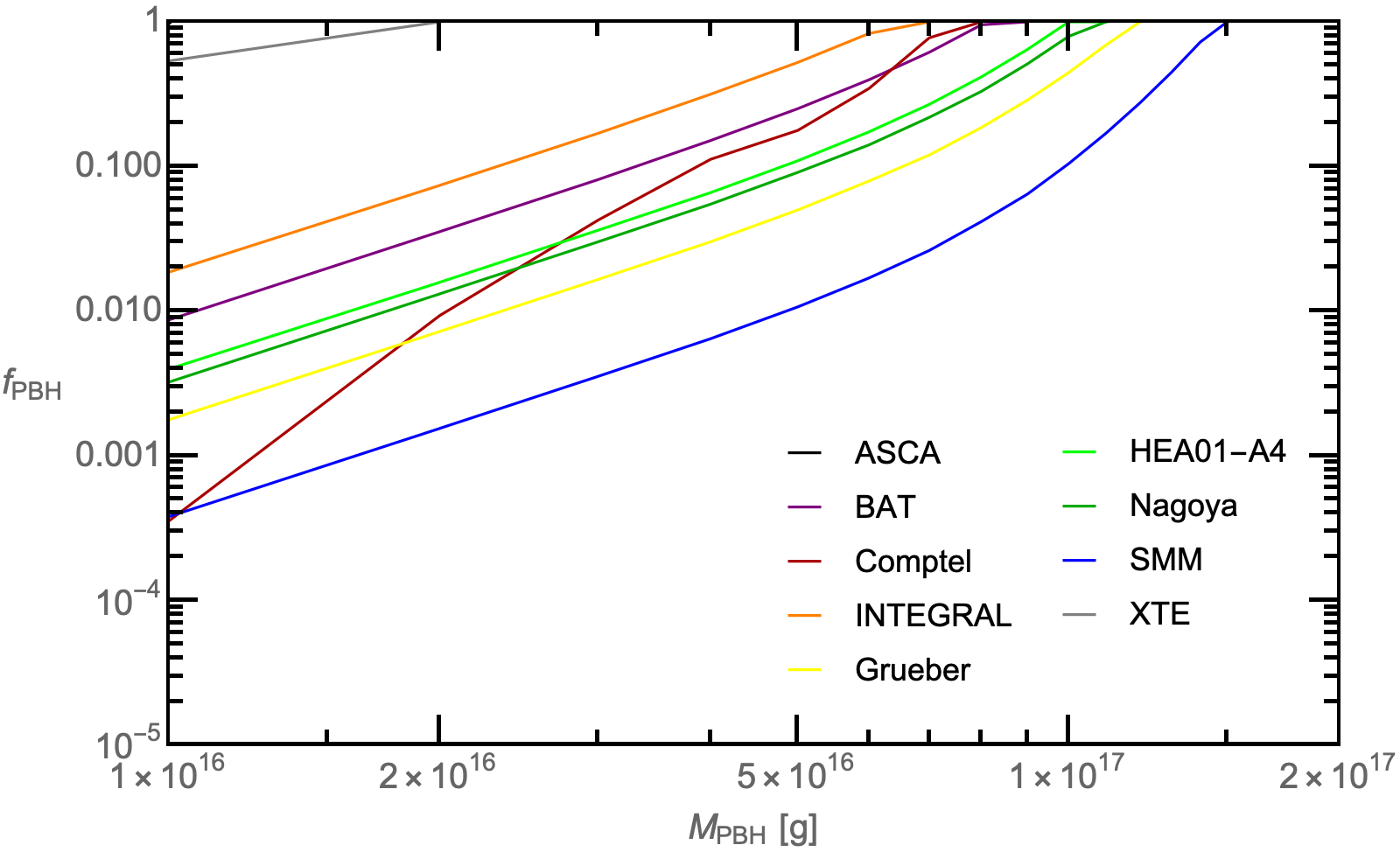}
    \caption{Bounds with no background modelling and $f_{\rm Ps}=1$ for the considered data sets.}
    \label{boundsfPs1noBGnew}
\end{figure}

First of all, we report in Fig.~\ref{boundsfPs0noBG} and Fig.~\ref{boundsfPs1noBGnew} the bounds derived for each dataset independently, with the ``no background'' approach described in the text and $f_{\rm Ps}=0$, $f_{\rm Ps}=1$, respectively.  

The trend is rather consistent with expectations, with higher energy data typically more constraining at lower PBH masses (for instance, Comptel is only relevant at $M\lesssim 2\times 10^{16}\,$g), and very low energy datasets such as ASCA and XTE poorly constraining, since no significant emission takes place in that band; some marginal constraint is nonetheless possible from these datasets if $f_{\rm Ps}=1$, since the low-energy tail is enhanced in that case, as shown in Fig.~\ref{isoflux} and in Fig.~\ref{totalflux}.

\begin{figure}[htbp!]
  \centering
    \includegraphics[width=0.45\textwidth]{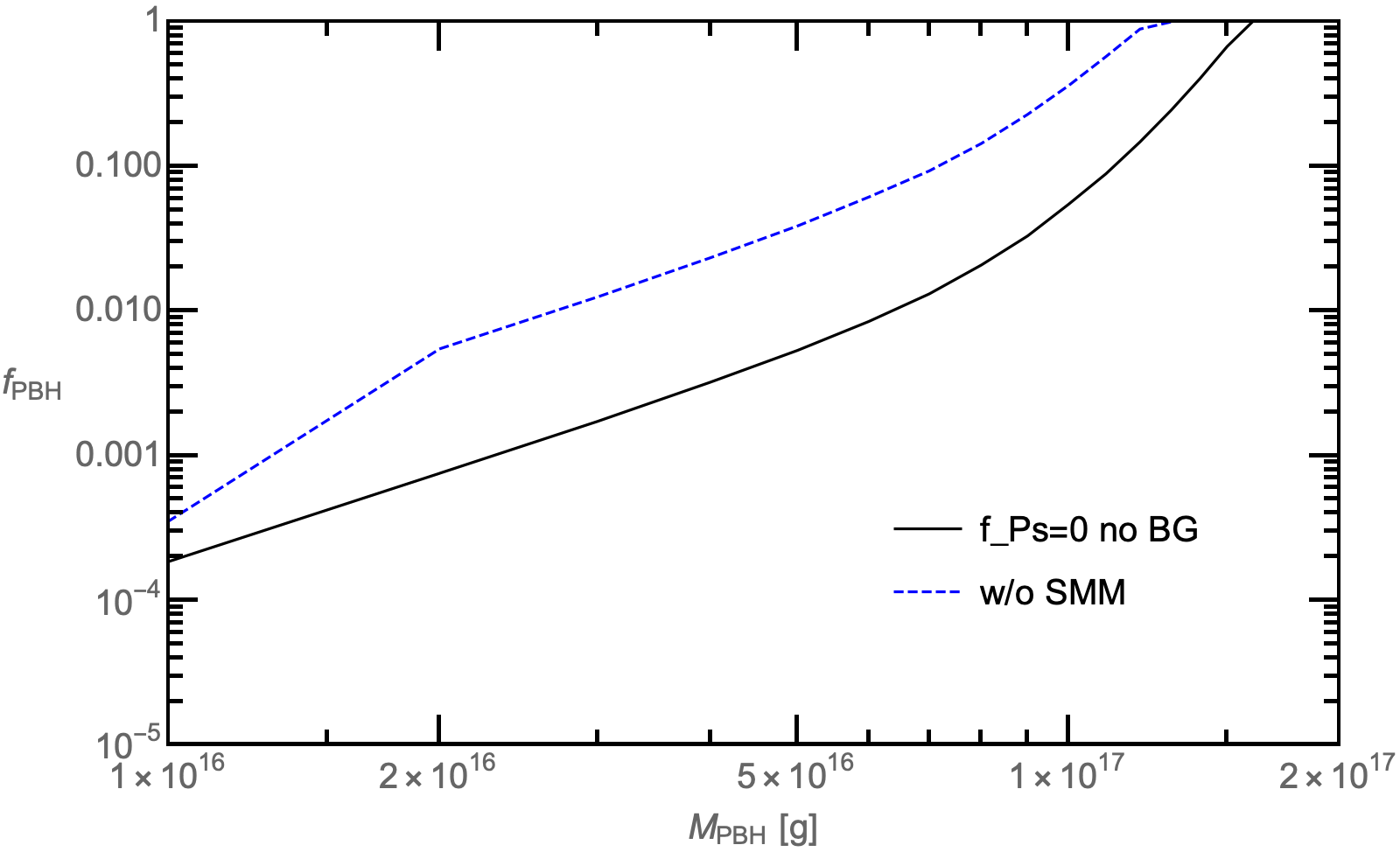}
    \caption{Bounds with no background modelling and $f_{\rm Ps}=0$ with (black solid) and without (blue dashed) SMM data.}
    \label{boundscombinedfPs0noBG}
\end{figure}

\begin{figure}[htbp!]
  \centering
    \includegraphics[width=0.45\textwidth]{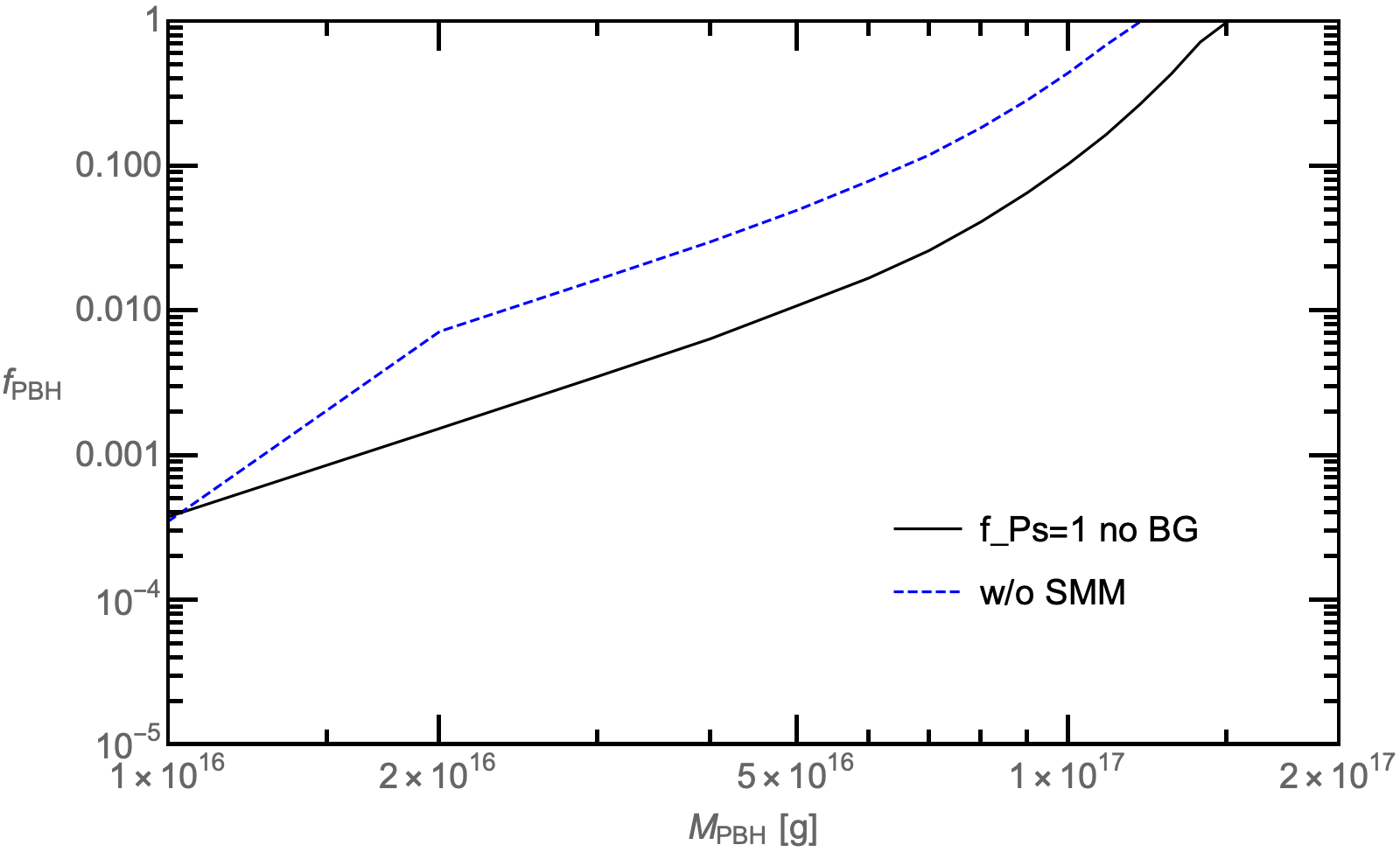}
    \caption{Bounds with no background modelling and $f_{\rm Ps}=1$ with (black solid) and without (blue dashed) SMM data.}
    \label{boundscombinedfPs1noBG}
\end{figure}

It is also interesting to check that the global constraint from the combination of all datasets is only mildly affected (i.e., a factor $\sim 2-3$) by the removal of the SMM dataset, which is the single ``most constraining'' dataset, as shown in Fig.~\ref{boundscombinedfPs0noBG} and Fig.~\ref{boundscombinedfPs1noBG} for $f_{\rm Ps}=0$ and $f_{\rm Ps}=1$, respectively. Even so, the bounds remain marginally stronger than the most competitive ones present in the literature. This is a gauge of robustness of the results with respect to unanticipated systematics in any of the observations included in our compilation. 

\begin{figure}[htbp!]
  \centering
    \includegraphics[width=0.45\textwidth]{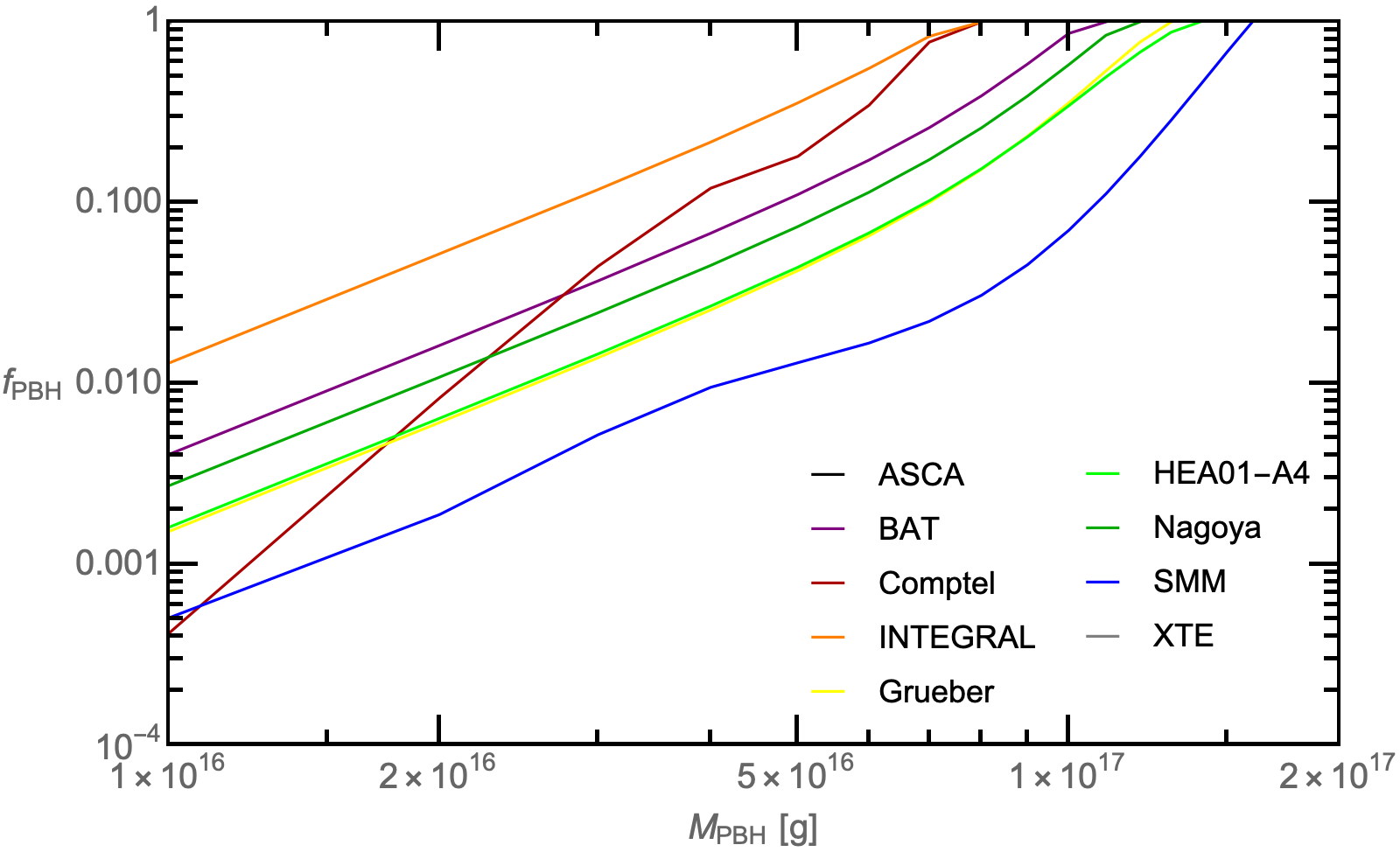}
    \caption{Bounds with background modelling and $f_{\rm Ps}=0$ for the considered data sets.}
    \label{boundsfPs0BG}
\end{figure}

\begin{figure}[htbp!]
  \centering
    \includegraphics[width=0.45\textwidth]{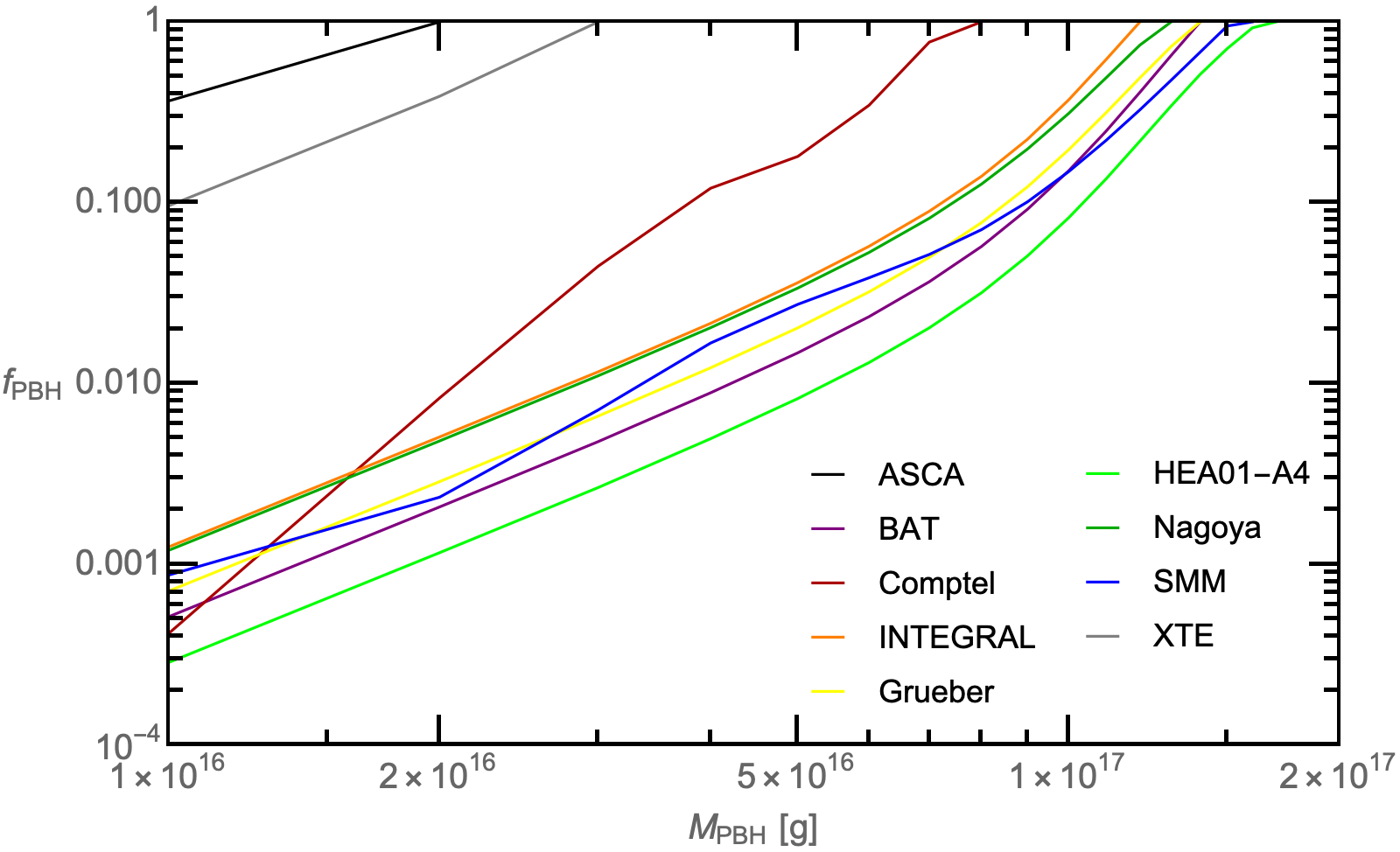}
    \caption{Bounds with background modelling and $f_{\rm Ps}=1$ for the considered data sets.}
    \label{boundsfPs1BG}
\end{figure}

We have repeated this exercise including the background model by Ueda et al., obtaining the corresponding single dataset bounds. We report our results in Fig.~\ref{boundsfPs0BG} and Fig.~\ref{boundsfPs1BG} for $f_{\rm Ps}=0$ and $f_{\rm Ps}=1$, respectively. 
While the overall trend is similar to the previous case, a notable difference is manifest in the role of the SMM dataset, which is less crucial for the  $f_{\rm Ps}=1$ case. This is the result of the lack of the prominent line at 511 keV in this circumstance. 
    
\begin{figure}[htbp!]
  \centering
    \includegraphics[width=0.45\textwidth]{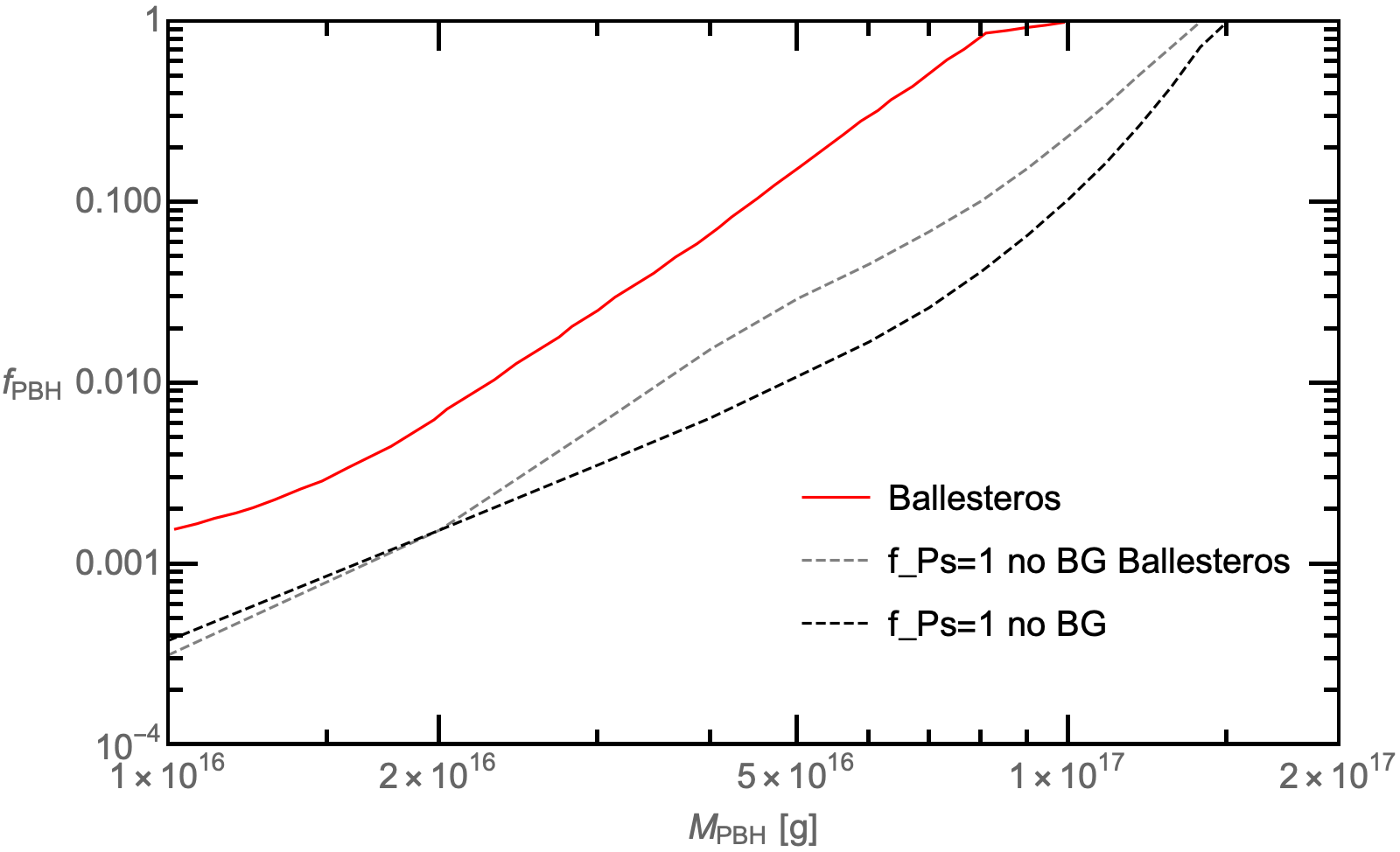}
    \caption{Bounds, for $f_{\rm Ps}=1$, derived following our conservative approach (no background modeling) described in the main text (black) and according to the estimator in Eq.~(\ref{estimatornoBG}), as done in~\cite{Ballesteros:2019exr} (gray). For comparison, the conservative bounds obtained in~~\cite{Ballesteros:2019exr} for the same datasets are reported.}
    \label{boundscombinedall}
\end{figure}

Finally, for the sake of comparison, we
also derive exclusion curves if adopting the 
same ``conservative'' approach without background as in \cite{Ballesteros:2019exr}:
For every pair of values for $\{ f_{\rm PBH}, M_{\rm PBH} \}$ in the range of interest, we compute the following estimator
\begin{equation}
 \chi^{2}=\sum^{O}_{i}\frac{\left(D_{i}-A_{i}\right)^{2}}{\Delta^{2}_{i}}\,,
\label{estimatornoBG}
\end{equation}
{\it limiting the sum to the $O$ bins for which the predicted flux $A_i$ overshoots the data}, where $D_{i}$ are the data points and $\Delta_{i}$ are the measurement uncertainties. Then, for a given mass $M_{\rm PBH}$, the maximum allowed value for $f_{\rm PBH}$ at 95 $\%$ C.L. is the one above which $\chi^{2} \le \chi^{2}_{0.05}(O-1)$ is not fulfilled anymore, where $\chi^{2}_{0.05}(O-1)$ is the 5\% value corresponding to $O-1$ degrees of freedom. Note that $O$ now depends on the PBH parameters, making the procedure more involved and statistically less transparent to assess than the one adopted in the main text. Nonetheless, we see that this approach leads to bounds very similar to the ones previously obtained at low or large masses, and within a factor $\sim 2$ of those in between. Clearly, the bulk of the
improvement over~\cite{Ballesteros:2019exr} is not attributed to the statistical procedure, but to the intrinsically larger theoretical fluxes we predict. 
\end{document}